\documentstyle[12pt]{article}
\newfont{\feff}{cmti10}
\newcommand{\newsection}{    
\setcounter{equation}{0}
\section}

\begin{document}

\begin{titlepage}
{\bf August, 1992}\hfill	  {\bf LPTENS-92/22}\\

\begin{center}

{\bf PHASE TRANSITIONS IN INDUCED QCD}

\vspace{1.5cm}

{\bf  A.A.~Migdal}\footnote{Permanent addres: Physics Department,  Princeton
University, Jadwin Hall,
Princeton, NJ 08544-1000. E-mail: migdal@acm.princeton.edu}

\vspace{1.0cm}

{\it Laboratoire de Physique Th\'eorique\footnote{Unit\'e propre du
CNRS, associ\'ee \`a l`Ecole Normale Sup\'erieure et \`a
l`Universit\'e de Paris Sud}\\
de L'Ecole Normale Sup\'erieure, 24 rue Lhomond,\\
75231 Paris CEDEX 05, France}

\vspace{1.9cm}
\end{center}

\abstract{
The variety of the phase transitions in Induced QCD are studied.
Depending upon the parameters in the scalar field potential, there
could be infinite number of fixed points, with different critical
behavior. The integral equation for the density of the eigenvalues of
the scalar field are generalized to the weak coupling phases, with the
gap at the origin.  We find a wide class of the massive solutions of
these integral equations in the strong coupling phases, and derive an
explicit eigenvalue equation for the scalar branch of the mass
spectrum.}
\vfill
\end{titlepage}


\newsection{Introduction}

Induced QCD was suggested  as a
possible model of hadrons two months ago ~\cite{KM92}. This
is the lattice model of the scalar
field $  \Phi $ in adjoint representation of the $ SU_{N} $ gauge group,
interacting with the usual lattice gauge field $ U_{l} $, defined at
links $l$. The unusual  feature, which allows one to solve the model exactly
in the large $ N $ limit ~\cite{Mig92a,Mig92b} for arbitrary dimension
$ D $ of the
lattice, is the absence of the bare self-interaction
for the gauge field.

The idea is, that this self-interaction would be induced  at larger
distances, where the scalar field decouples, being heavy. Such induction
often takes place in two-dimensional models of gauge and gravitational
fields, such
as $ \mbox{CP}_N $ models, or the Liouville theory.

The parameters of the scalar potential $ U(\Phi) $ should be carefully
adjusted for this miracle to occur in more than two dimensions. The
bare mass $
m_0^2 = U''(0) $  should be chosen so,
that the effective scalar mass $ m_{eff}^2 $ becomes much less than the lattice
cutoff. This
effective mass $ m_{eff}$ would serve as the ultraviolet cutoff for
perturbative QCD.

At the scales less than  this mass, the scalar
field could be integrated out,  which  yields the
Yang-Mills term $ N \beta \mbox{ tr} F_{\mu,\nu}^2 $ in effective
Action, along with a variety of unwanted higher order terms. The bare
quartic coupling should tend to a hypothetical critical point, to provide
a large positive coefficient $ \beta $,
which could serve as the bare coupling of perturbative QCD. Then the
physical mass scale $ m_{phys} $  would automatically come out  $
m_{phys} \sim m_{eff} \exp \left(- A \beta \right)\, \beta^{B} $, with
coefficients $ A,B $ known from perturbative QCD.

The computation of induced gauge coupling $ \beta $ as a function of the
bare scalar mass and the bare quartic coupling $ \lambda_0 = U''''(0)
$ is not an easy task. Apparently, the quartic coupling should tend to
some ultraviolet fixed point of the renormalization group,
corresponding to both gauge and scalar charges. The recent computer
simulations ~\cite{GS92} for the $ SU_2 $ model showed the line of the
first order phase transitions in the $ m_0^2,\lambda_0$ plane, with the
endpoint, which might correspond to the second order phase
transition.

The general equations of ~\cite{Mig92a} were studied in great detail
for for special cases $D=1$  and quadratic potential
$U(\phi) $~\cite{Gr92}. In agreement with computer simulations and theoretical
expectations, there  is no critical point of the QCD type for the
quadratic potential. At $ D=1 $ in continuum limit the familiar
fermionic solution was reproduced. The more general solution, for the
finite lattice spacing at $D=1$ was found in ~\cite{CAP92}, again in
complete agreement with previous work on matrix models.

There are some lattice artifacts in this model, which it is not yet
clear how to remove. Namely, there is an extra local $ Z_N $ symmetry,
which must break to allow quarks to move inside hadrons. The general
mechanisms of this symmetry breaking were found long ago by
Khokhlachev and Makeenko~\cite{KhM81}, and later confirmed by numerical
experiments and mean field analysis ~\cite{MC,Monopole,Sam83,MF}.

It was recently conjectured in ~\cite{KSW92}, that the same mechanisms
spontaneously break $ Z_N $ symmetry in this model at the critical
point, which corresponds to the continuum limit of the lattice theory.
In a subsequent paper ~\cite{KhM92} it was argued on the basis of the
mean field analysis, that the $Z_N$ transition must take place before
the critical point, i.e. still in the strong coupling phase of our
model, if it really induces QCD.

Various properties of the generalized Wilson loops in this model were
studied in~\cite{KMSW92}. Some interesting mathematical structures
were found, which could be used in gauge and string models regardless
conjectured induction of QCD.

There is the general heuristic argument ~\cite{Mig92a}, that the nontrivial
fixed point of this model could be nothing but  QCD. The
argument is based on the common belief, that there is only one
nontrivial theory in four dimensions: the asymptotically free, quark
confining QCD. Even if this argument fails, the confining solution of
Induced QCD would be an exciting alternative to the usual QCD. This
would be the first solvable model of the QFT in four dimensions.

In the second paper ~\cite{Mig92a} we came very close to this
goal, by reducing the solution of the model to the following
nonlinear integral equation for the vacuum density $ \rho(\lambda) $ of the
scalar field eigenvalues
\begin{equation}
\wp \int_{-\infty}^{+\infty}\, d  \lambda' \left(
\frac{\pi\rho(\lambda')}{ \lambda'-\lambda}+\arctan \frac{\pi
\rho(\lambda')}{\lambda-R(\lambda')} \right) =0,
\label{MFE}
\end{equation}
where
\begin{equation}
  R(\lambda) = \frac{1}{2D}U'(\lambda) + \frac{D-1}{D} \wp
\int_{-\infty}^{\infty} \,d \lambda'
\frac{\rho(\lambda')}{ \lambda-\lambda'},
\end{equation}
and $ U(\phi) $ is the scalar field potential. We call this equation the master
field equation, or MFE, because this density could be regarded as the
long-sought master field of QCD.
We found exact powerlike solution at $ \lambda \rightarrow 0
$\footnote{The eigenvalues $ \lambda $ have dimension $ m^{\frac{1}{2}
D -1} $ so that at $ D>2 $ the physical region is $ \lambda
\rightarrow 0 $ in the lattice units we are using.}
\begin{equation}
  \rho(\lambda) \propto |\lambda|^{\alpha}\\;\; \cos \pi \alpha =
-\frac{D}{3D-2}\\; \; \alpha >1,
\end{equation}
which was quite encouraging, since the scaling index $ \alpha $ showed
no pathologies, like those of the string models.

The forbidden interval here is $ \frac{1}{2} < D < 1 $, and the only
rational values are $ \alpha = n+\frac{1}{2} $ at $ D=0 $, $ \alpha =
2n $ at $ D=\frac{1}{2} $, $ \alpha = 2n+1$ at $ D=1 $ and $ \alpha =
2n+1 \pm \frac{1}{3} $ at $ D=2 $. The solutions at $ D=0, \frac{1}{2}
$ are unphysical, as the assumption of vanishing density at the origin
is never satisfied in the corresponding matrix models.  $D=0$
corresponds to the one matrix model, and $ D=\frac{1}{2} $ correspond
to the two matrix model (the number $2D$ of links meeting in each cite
equals $ 1 $ here, as there is only one link, connecting two cites).
$D=1$ solution is already physical, as we discuss in more detail
below. The $ |\lambda|^{2n+1} $ singularity comes about as the
singularity of the tip of the upside-down even potential in the usual
solution of the $D=1$ matrix models.  As for the first nontrivial case
$D=2$, unfortunately, the adjoint scalar field model cannot be solved
by conventional methods even at $ D=2 $, so there is nothing to
compare with this solution.

The $  \frac{1}{ N} $ expansion was
considered in the third paper ~\cite{Mig92b}, where we found the integral
equation for the  propagator of the effective field theory with $
\rho(\lambda,x) $ as dynamical field. This linear equation involves
the vacuum density in its kernel, and for the powerlike density, the
powerlike solutions for the corresponding wave functions in $ \lambda
$ space were found.

Still, the solution is incomplete, as there is no mass scale. This
is the solution exactly at the critical point, where there are scaling
laws in the $ \lambda $ space. \footnote{As was discussed already in the
first paper, the renormalization group analysis tells us, that the
logarithmic laws of the asymptotic freedom translate into the power
laws in the induced QCD models. The critical indices depend upon the
effective quartic scalar interaction, which is not calculable by
perturbative methods.}
The physical solution of Induced QCD must involve the mass scale, which
requires the more general solution of the MFE.

In this paper we find an infinite family of such solutions, which
turns out to be a particular superposition of the previous powerlike
terms.
The implications of this simple observation are very
interesting. Now, there is a  calculable mass spectrum with nontrivial
scaling indices in arbitrary dimension $ D>1 $.

We generalize the Riemann-Hilbert problem
for the weak coupling phase, where there is a gap at the origin in the
density of eigenvalues. We present the new derivation, which, as we hope, is
easier to comprehend, than that of ~\cite{Mig92a}.

\newsection{Massive Solution of the  Riemann-Hilbert Problem}

The classical equation (\ref{MFE}) in the local limit, when $
r(\lambda) \equiv R(\lambda) - \lambda \sim \rho(\lambda) \ll \lambda $  was
reduced in the previous paper ~\cite{Mig92a} to the following nonlinear
boundary problem.

Let us introduce two   functions
\begin{equation}
  {\cal P}(z) = \frac{U'(z)-2D z}{2(1-D)}+ \int_{- \infty}^{ \infty} d\mu
\,\frac{\rho(\mu)}{\mu-z},
\end{equation}
\begin{equation}
  {\cal Q}(z) =  \mbox{polynomial}
+ \pi \int_{- \infty}^{ \infty} d\mu \,\frac{\rho^2(\mu)}{\mu-z}.
\end{equation}
At $  z \rightarrow \lambda +\imath 0 $ , we have
\begin{equation}
  {\cal P}(z) \rightarrow \frac{D}{1-D} r(\lambda) + \imath  \pi \rho(\lambda).
\end{equation}
Here $ U(\phi) $ is the bare
potential, but the contributions from the large eigenvalues ( of the
order of the lattice cutoff) in the integral effectively renormalize
this potential. In the local limit, at $ \lambda \ll 1 $ in lattice
units, the density  $ \rho \sim |\lambda|^{\alpha} $, so that formally
the integral
diverges if we substitute the local density, i.e., it is
dominated by the lattice scales $ \lambda \sim 1 $, where the solution
is not universal. The corresponding number of subtractions should be made, as
usual in dispersion relations, or, which is more convenient, one could
make the analytic continuation of these integrals in the scaling dimension $
\alpha $ from the convergence region $ \alpha < \frac{1}{2}  $. The subtraction
polynomial renormalize the bare potential $ U $.

Both functions are analytic in the upper half plane and have the
symmetry property
\begin{equation}
  {\cal P}(-\bar{z}) = -\bar{{\cal P}}(z)\\;\;
  {\cal Q}(-\bar{z}) = -\bar{{\cal Q}}(z).
\label{RH1}
\end{equation}
In other words, real(imaginary) parts are odd(even) with respect to
the real part of $ z $.

At $ \Im z \rightarrow +0$ by construction
\begin{equation}
  \Im\,{\cal Q} =  \left(\Im{\cal P}\right)^2.
\label{RH2}
\end{equation}
On the other hand, as shown in  ~\cite{Mig92a}, in virtue of the  classical
equation  (\ref{MFE}), up to $ O({\cal P}^3) $ terms\footnote{These terms
are down by a power of the ultraviolet cutoff in the local limit.}
\begin{equation}
  \Re {\cal Q} = \frac{1-D}{D} \Im\left({\cal P}^2\right).
\label{RH3}
\end{equation}
The last three equations represent  the nonlinear Riemann-Hilbert
problem.

In the previous papers only the massless solutions were found
\begin{equation}
  {\cal P} = \imath A \frac{\left(-\imath z \right)^{\alpha}}{\cos
\frac{\pi \alpha}{2}},
\end{equation}
\begin{equation}
  {\cal Q} = \imath  A^2  \frac{\left(-\imath z
\right)^{2\alpha}}{\cos \pi \alpha}.
\label{QP}
\end{equation}
Various values of $ \alpha = 2n +1 \pm \frac{1}{\pi}
\arccos\frac{D}{3D-2} $ correspond to various fixed points of the model.

We could not find any general theory of the nonlinear Riemann-Hilbert
problem, but in this particular one we found  the class of
exact solutions, which are  built from the above power terms,
\begin{equation}
  {\cal P}=  \imath A \frac{\left(-\imath z \right)^{\alpha}}{\cos
\frac{\pi \alpha}{2}}
+ \imath B \frac{\left(-\imath z \right)^{\beta }}{\cos
\frac{\pi \beta }{2}}\\;\; \alpha+ \beta = 2 k,
\label{MASSP}
\end{equation}
\begin{equation}
  {\cal Q} = \imath \frac{ A^2 \left(-\imath z\right)^{2\alpha}
+ B^2 \left(-\imath z\right)^{2 \beta}}{\cos \pi \alpha} + 2 \imath A
B z^{\alpha + \beta}
\label{MASSQ}.
\end{equation}

It is not difficult to check this solution. The symmetry property is
manifest, and so is the first equation. As for the second equation,
the key point is that at $ z= \lambda + \imath 0 $
\begin{equation}
 \Im {\cal P}= A |\lambda|^{\alpha} + B |\lambda|^{\beta}\\;\;
 \Re {\cal P}= \frac{\lambda}{|\lambda|} \tan \frac{\pi
\alpha}{2}\left(A |\lambda|^{\alpha} - B |\lambda|^{\beta}\right),
\end{equation}
so that the cross terms in $ \Im ({\cal P}^2) = 2 \Im {\cal P} \,
\Re {\cal P} $ are absent.

In general, these $ A $ and $ B $ could be arbitrary polynomials of $
z^2 $. This ambiguity reflects the ambiguity in the choice of the
initial scalar potential $ U(\phi) $. However, the critical phenomena,
which we are interested in, are universal, as they take place in the
infinitesimal vicinity of the origin, in the lattice units we are
using.

In this limit, only the two leading terms in $ A, B $ can be left. With
proper redefinition of $ \alpha, \beta $ this corresponds to constant
$ A,B $. We assume, that $ \alpha > \beta $, then the critical region
corresponds to
\begin{equation}
  z \sim z_0 \equiv \left( \frac{B}{A} \right)^{\omega } \\;\; \omega
= \frac{1}{ \alpha- \beta},
\end{equation}
so, that $ B=0 $ at the critical point. The leading index $ \alpha $ must be
greater than $ 1 $, as it follows from original derivation. As for the
subleading index $ \beta $, it should be greater than $ -1 $ for the
density to be integrable at the origin.

Using the language of renormalization group, the $ \beta $-term
represents the perturbation of the UV-stable fixed point $ \alpha $ by
the relevant operator of the lower scaling dimension. In general, the
coefficient $ B $ linearly vanishes at the critical point. At $ B< 0 $
the solution becomes unstable, as the density changes sign near the
origin. In this case, there would be the phase transition to the weak
coupling phase. The local limit of the MFE is different in this phase,
as we shall see in the next Section.

The lowest  solution for $ \alpha, \beta $ in the strong
coupling phase would be $ 1 < \alpha  < 1.5$,
and $ \beta   = 2-\alpha  $. In four dimensions
\begin{equation}
  \{\alpha,\beta\}  = \{1.36901,0.63099 \}.
\label{CR2}
\end{equation}
The next one is
\begin{equation}
  \{\alpha,\beta\}  = \{2.63099,-0.63099\}.
\end{equation}
Note that the singularity in $ \rho $ is still integrable for this
solution. These are the only physical solutions for $ k=1$. At the
next level, $ k=2 $, there are three solutions
\begin{equation}
  \{\alpha,\beta\}  = \{\{2.63099, 1.36901\}, \{3.36901, 0.63099\},
\{4.63099, -0.63099\}\}.
\end{equation}
We discuss the choice of the solution later, when
we study the wave equation. As we shall see, the first  solution at $
k=1 $ have tachyons.

It is worth mentioning, that at $ D=1 $ the equations degenerate. The
real part of $ {\cal Q} $ vanishes, so that the most general solution
with proper symmetry would be $ \imath W(z^2) $, where $W $ is some
polynomial with real coefficients. Then, from the first equation we
find
\begin{equation}
  \pi \rho(\lambda) = \sqrt{W(\lambda^2)}.
\end{equation}

This is in complete agreement with the solution of the lattice MFE, found
recently by D.Gross ~\cite{Gr92}. In this case $ W(z^2) =2 \pi^2(E- U(z)) $
where $ U(z) $ is initial potential, and the chemical potential $ E $
 is to be determined from the normalization of
density. This solution also agrees with conventional solution of the $
D=1 $ model in terms of  effective fermi gas. Note, that the generic
singularity is $ |\lambda|^{2n+1} $, for $ W(\lambda) \propto
\lambda^{4n+2} $. This agrees with above powerlike solution.

With our "S-matrix" approach we immediately find the solution, but
cannot relate the parameters to those of original lattice theory.
Fortunately, this is never needed. What is really needed, is to check
internal consistency of the solution, such as absence of ghosts and
tachyons, which is not apriori guaranteed in the "S-matrix" approach.

Another comment. At any $ D $ there always exists a trivial  solution, without
critical behavior
\begin{equation}
  {\cal P} = z a(z^2) + \imath \sqrt{b(z^2)}\\;\; {\cal Q} = \imath
b^2(z^2) + 2 \frac{1-D}{D}z a(z^2) \sqrt{b(z^2)},
\end{equation}
with the support of eigenvalues at $ b(\lambda^2) >0 $. In the simplest
case of constant $ a $ and linear $ b $ this is the semicircle
solution. As was recently shown in ~\cite{Gr92}, for the case of
quadratic potential $ U $ this is the only solution in the strong
coupling phase.

Unfortunately, the simplest nontrivial scaling solution would take at
least two adjustable parameters, so the higher order terms in a
potential are required. In this case, as we suspect, the explicit
solution of the lattice model is unavailable, and one either has to rely upon
the above Riemann-Hilbert approach, or use the numerical methods
to solve the lattice MFE.

\newsection{Phase Transition}

The Riemann-Hilbert problem was derived under assumption, that there
was infinite support of the eigenvalues, without any gap at the
origin. As was mentioned in the first paper ~\cite{KM92}, we expect
this model to undergo the phase transition from above strong coupling
phase to the weak coupling phase, with the gap in the support of the
eigenvalues.  In that paper we could not find nontrivial spectrum,
because the kinetic term in the effective Lagrangean for the density
fluctuations vanished at $ N = \infty $.

The more recent solution ~\cite{Mig92b} produces such term regardless
the phase of the model. The term proves to be positive definite, which
means that this solution is different from the first one. Perhaps,
there was something wrong with the assumptions of the orthogonal
polynomial method in this case.\footnote{ In some cases, the even and
odd coefficients of expansion in orthogonal polynomials tend to
different analytic functions of $ \frac{n}{N} $.  Certainly, this
issue must be further analyzed.}

Anyway, let us assume, that there is a gap from $ -a $ to $ a $ in the
vacuum density $ \rho(\lambda)$. The basic equation (\ref{MFE})
remains the same, but the dispersion relation between real and
imaginary parts of all analytic functions modifies. The simplest way
to account for these changes is to note that conformal transformation
\begin{equation}
  \zeta(z) = \sqrt{z^2 -a^2}
\end{equation}
maps the  upper half of $ z $ plane onto the upper half plane of $
\zeta $, eliminating the gap from $ -a $ to $ a $.

We could use old dispersion relations {\em for the
even functions of z} with $ \zeta $ instead of $ z $. For the odd
functions there would be fictitious singularity at $ z^2 = 0 $.  In
particular, the odd function $ {\cal P}(z) $ was reconstructed in
{}~\cite{Mig92a}
from the real part. The  derivative of this relation reads
(so far, at $ a=0$),
\begin{equation}
 {\cal P}'(z) =  \frac{U''(z)-2D}{2(1-D)} +
\imath\int_{-\infty}^{\infty} \frac{d \nu}{\pi}
\left(  \frac{1}{z -C(\nu)} -  \frac{1}{
z -R(\nu)} \right),
\end{equation}
where
\begin{equation}
C(\nu) \equiv R(\nu) -\imath \pi \rho(\nu).
\end{equation}
This is the difference of two Cauchy integrals, the first one going
over the complex curve
$ z = C(\nu) $ in the lower half plane, and the second one going
backwards over the
real axis $ z=R(\nu) $. One could write this as a single Cauchy
integral over the complex contour $ {\cal C} =\{C,R\}$,
\begin{equation}
  {\cal P}'(z) =  \frac{U''(z)-2D}{2(1-D)} +
\imath \oint_{ {\cal C}}\frac{d y}{\pi} \frac{ W_{ {\cal C}} (y)}{z- y},
\end{equation}
where the density $ W_{ {\cal C}} $ is given by the parametric equation
\begin{equation}
  W_{ {\cal C}}\left({\cal C}(\nu) \right) =  \frac{1}{ {\cal C}'(\nu) }
\end{equation}

Transforming this relation from $ z $ to $ \zeta $, we  find
\begin{equation}
  {\cal P}'(z) =  \frac{U''(z)-2D}{2(1-D)} +
\imath \oint_{ {\cal C}}\frac{d y}{\pi} \frac{ W_{ {\cal C}}
(y)}{\zeta(z)- y},
\label{GAP}
\end{equation}
\begin{equation}
  W_{ {\cal C}}\left(\zeta \left({\cal C}(\nu) \right) \right)  =
\frac{1}{ {\cal C}'(\nu) }
\end{equation}
In terms of original variables
\begin{equation}
  {\cal P}'(z) =  \frac{U''(z)-2D}{2(1-D)} +
\imath\int_{-\infty}^{\infty} \frac{d \nu}{\pi}
\left( \frac{\zeta'(C(\nu))}{\zeta(z)
-\zeta(C(\nu))} -
\frac{\zeta'(R(\nu))}{\zeta(z)
-\zeta(R(\nu))}
\right)
\end{equation}
Note, that the integrand vanishes inside the gap, as $ C(\nu)=R(\nu) $
there. Also, note, that in virtue of the
symmetry $ R(-\nu) = - R(\nu), \rho(-\nu) = \rho(\nu) $ the real
(imaginary) part of this function is odd(even) function of $ \Re z $,
as it should be.
One can readily check that this function is analytic in the upper half
plane, that its imaginary part at the real axis vanishes at $ z^2 <
a^2 $, and that its real part  at $ z^2 > a^2 $ agrees with derivative
of MFE. \footnote{The sceptical reader is invited to
check this formula by means of usual dispersion relations in the $ z^2 $
plane, with extra factors $ \sqrt{a^2-z^2} $ to convert real part to
imaginary. In differentiating the MFE, one should take into account
the $ \delta $-function terms, coming from the discontinuity of the
actangent at $ \pm \infty $.}

Let us now go to the local limit in above complex contour integral for
$ {\cal P}'(z) $, using the same expansion, as in  ref. ~\cite{Mig92a},
\begin{equation}
  R(\nu) =\nu + r(\nu)\\;\; C(\nu) = \nu+c(\nu)\\;\;
c(y) = r(y) - \imath \pi \rho(y) \ll \nu.
\end{equation}
Iterating implicit equation for $ W_{ {\cal C}} $, we find
\begin{equation}
  W_C(\zeta(z)) =  1 -c'(z) + \frac{1}{2} \left(c^2(z)\right)''+ O(c^3)
\end{equation}

We are interested in $\pi \rho'(\lambda ) = \Im  {\cal P}'(\lambda+
\imath 0) $. One contribution to this imaginary part  comes from
$ \Im  W(\zeta(\lambda)) $ times the $ \delta $-function term at $ y=
\zeta(\lambda)  $.
Subtracting the same terms with $ \rho=0 $, to account for
the integral with $ C(\nu) $ replaced by $ R(\nu) $ , we find
\begin{equation}
\Im  W_{C}(\zeta(\lambda)) - \Im  W_R(\zeta(\lambda)) =
\pi \rho'(\lambda) -\pi\left(r(\lambda)\rho(\lambda)\right)''
\end{equation}
We see, that the left side exactly cancels with the first term, so
that we have to keep the $ O(c^2) $ terms.

There is the second contribution to $\Im  {\cal P}'(\lambda+
\imath 0) $, coming from the
principal value of the integral of $  \Re  W_{{\cal C}} $. As before,
subtracting the integrals with $ C $ and $ R $ , we find
\begin{equation}
 -\frac{\pi}{2}\,  \wp\,\int_{-\infty}^{\infty}  dy \,
\frac{1}{(\sqrt{\lambda^2-a^2}  -y)}
\left(\rho^2(\nu)\right)''_{\nu= \sqrt{a^2+y^2}},
\end{equation}
or, symmetrizing in $ y \rightarrow -y  $ and transforming $ y d y = \nu d \nu
$,
\begin{eqnarray}
 &&-\pi\sqrt{\lambda^2-a^2}\,  \wp\,\int_{a}^{\infty}
\frac{d\nu}{\sqrt{\nu^2-a^2}} \, \frac{\nu}{(\lambda^2-\nu^2)}
\left(\rho^2(\nu)\right)''= \\ \nonumber & \,&
-\frac{\pi}{2} \sqrt{1-\frac{a^2}{\lambda^2}}\,  \wp\,\int_{S}
\frac{d\nu}{\sqrt{1-\frac{a^2}{\nu^2}}} \, \frac{1}{(\lambda-\nu)}
\left(\rho^2(\nu)\right)''
\end{eqnarray}
where $ S =\{(-\infty,-a),(a,\infty)\}$ is the support of eigenvalues.

Collecting the terms,  we arrive at the following equation
\begin{equation}
 2\left( r(\lambda) \rho(\lambda) \right)''
=  -\sqrt{1-\frac{a^2}{\lambda^2}}\, \wp\,\int_{S}
\frac{d\nu}{\sqrt{1-\frac{a^2}{\nu^2}}}
\frac{\left(\rho^2(\nu) \right)'' }
{\lambda-\nu}
\label{LMFEA}
\end{equation}
At $ a=0 $ this equation reduces to the old one, after two
integrations by parts.  We drop the divergent polynomial terms,
keeping in mind the corresponding number of subtractions
in dispersion relations.

The dispersion relation for $ r(\lambda) $ reads, as before
\begin{equation}
  r(\lambda) = \frac{U'(\lambda)-2 D \lambda}{2 D} + \frac{D-1}{D}
\, \wp \int_{S} \, d \nu \frac{\rho(\nu)}{\lambda-\nu}.
\end{equation}

Let us now reduce these equations to the Riemann-Hilbert problem. The
first function, $ {\cal P}(z) $ is the same as before. The second
function, $ {\cal Q}(z) $ is introduced as follows
\begin{equation}
  {\cal Q}''(z) =  \pi  \int_{S}
\frac{d\nu}{\sqrt{1-\frac{a^2}{\nu^2}}}
\frac{\left(\rho^2(\nu) \right)'' }{\nu-z}
\end{equation}

This function has the same symmetry and analyticity properties, as the
first one, including the gap in imaginary part at $ z=
\lambda+\imath 0 $
\begin{equation}
 \sqrt{1-\frac{a^2}{\lambda^2}}\, \Im  {\cal Q}'' =
\theta\left(\lambda^2-a^2 \right)
\left[(\Im  {\cal P})^2 \right]''
\end{equation}
As for the real parts, they are related at $\lambda^2 > a^2 $ as follows
\begin{equation}
\sqrt{1-\frac{a^2}{\lambda^2}} \,\Re  {\cal Q}'' = \frac{1-D}{D}
\left[\Im {\cal P}^2 \right]''
\end{equation}

At finite gap, we cannot eliminate the derivatives, because we cannot
include the factor $\sqrt{1-\frac{a^2}{z^2}}$ in $ {\cal Q}''(z) $
without introducing the singularity at $ z=0 $.
Still, the equations are so simple and universal, that one may hope to
find the analytic solution, like the one we found in the strong
coupling phase.

\newsection{The Mass Spectrum in the Strong Coupling Phase}

Let us now substitute the above general solution (\ref{MASSP})
into the  wave equation, found in the previous paper ~\cite{Mig92b}. In present
notations, with
\begin{equation}
  M^2_{eff}(z) \equiv \frac{ D U''(z)-2D^2 -2 D +2}{D-1} +
\mbox{polynomial} = \tau_0 + \tau_1 z^2 + \dots,
\end{equation}
the wave equation reads (at $ z = \lambda + \imath 0 $)
\begin{equation}
 -\frac{1}{2} \left(P^2+ M^2_{eff}  \right)\,\Im {\cal F}=
D \,\Re {\cal P}' \,\Im {\cal F} - \Im {\cal P}\, \Re{\cal F}' +
\frac{(D-1)^2}{D} \Im  {\cal P}
\left(\frac{ \Re {\cal P}\, \Im  {\cal F}}{\Im  {\cal P}} \right)',
\end{equation}
where $ P $ is the Euclidean 4-momentum of the vacuum excitation,
corresponding to plane wave fluctuations of $ \rho $,
\begin{equation}
\delta \rho(\lambda,x) =  \frac{1}{ N} e^{\imath P x}
 \wp \int_{-\infty}^{\infty} \, d\nu \frac{\Im {\cal F}(\nu+
\imath0)}{\pi^3\rho(\nu)}
  \frac{1}{\nu-\lambda}.
\end{equation}

The $ \frac{\lambda}{|\lambda|} $ term in the ratio of real and
imaginary parts of $ {\cal P} $ yields the $ \delta(\lambda ) $ term
which drops provided
\begin{equation}
  \Im  {\cal P}\, \Im {\cal F} = 0 \mbox{ at $ z =0 $}.
\label{BCON}
\end{equation}
This boundary condition selects the physical
solutions.

In the simplest nontrivial case of the mass term plus quartic
interaction $ U(\lambda) = \frac{1}{2}  m_0^2 + \frac{1}{4}  \lambda_0
\lambda^4 $ there are two terms $  \tau_0, \tau_1 $ present in
effective mass term. The first term $ \tau_0 $ must vanish as $
z_0^{\alpha-1} $ to be relevant in the critical region $ z \sim  z_0 $.
This yields the equation
\begin{equation}
  \tau_0 \propto B^{\delta_0 } \\;\;
\delta_0 =\frac{\alpha-1}{\alpha-\beta}.
\end{equation}
In terms of
the original parameters of the scalar potential, above relation  describes the
curve of the first order phase transitions in the $ m_0^2, \lambda_0 $
plane, ending at the critical point, where $ \tau_0 = B = 0$.
This is is qualitative agreement with the simulations of ~\cite{GS92}.

When the $ \alpha >3 $ solution is taken, the first $ \tau_1 $
correction to $ M_{eff}^2 $ is relevant. The similar estimate yields
the scaling relation
\begin{equation}
  \tau_1 \propto B^{\delta_1 } \\;\;
\delta_1 =\frac{\alpha-3}{\alpha-\beta}.
\end{equation}
This would correspond to the tricritical point.
In general, for $ \alpha > 2 m + 1 $ the $ \tau_m z^{2m} $ terms in $
M^2_{eff}$, coming from the $ \phi^{2m+2} $ terms in $ U(\phi) $,
become relevant,
\begin{equation}
  \tau_m \propto B^{\delta_m } \\;\;
\delta_m =\frac{\alpha-2m-1}{\alpha-\beta}.
\end{equation}
%
%

Let us denote
\begin{equation}
  \alpha = k + \mu \\;\; \beta = k-\mu\\;\; \mu>0.
\end{equation}
Note that the mass indexes
\begin{equation}
   \delta_m=\frac{1}{2}+\frac{k-2m-1}{\mu},
\end{equation}
are trivial only for the simplest fixed point, with $ k=1, m=0 $. In
general, these are transcendental numbers, which agrees with the
induced QCD scenario, and contradicts the Gaussian fixed point for the
scalar field.

Consider  the infinite sum of  power terms for $ {\cal F} $
\begin{equation}
  {\cal F} = \imath^s\sum_{\epsilon}\, \frac{f(\epsilon) \left(-\imath z
\right)^{\epsilon}  }
{\sin \frac{\pi (s-\epsilon)}{2}},
\end{equation}
where $ s= \{ 0,1\} $ is the "$ \lambda $-parity"
\begin{equation}
  {\cal F}(-\bar{z}) = (-1)^s \bar{ {\cal F}}(z) \\;\; \delta
\rho(-\lambda,x) = (-1)^s \delta \rho(\lambda,x).
\end{equation}
Differentiating real and imaginary parts of $ {\cal P} , {\cal F} $,
multiplying by $ \Im  {\cal P} $ and collecting  the power terms, we
find the following equation
\begin{eqnarray}
 && \sum_{\epsilon}
f\left(\epsilon\right)\lambda^{\epsilon}
\left[-\frac{1}{2} (P^2+M^2_{eff}(\lambda))\left(A \lambda^{\mu}+B
\lambda^{-\mu}\right)\right] \\ \nonumber & \,&
= \sum_{\epsilon}
f\left(\epsilon\right)\lambda^{\epsilon+k-1}
\left[
A^2\Phi_2(\epsilon) \lambda^{2\mu}+ 2 A B \Phi_0(\epsilon) +
B^2 \Phi_{-2}(\epsilon) \lambda^{-2\mu} \right],
\end{eqnarray}
where
\begin{eqnarray}
&& M^2_{eff}(\lambda)= \sum_{m=0}^{\lfloor \frac{1}{2}(k+\mu-1)
\rfloor}\tau_m \lambda^{2m} \\ \nonumber & \,&
\Phi_{\pm2}(\epsilon)= \epsilon  \cot
\left(\frac{\pi (\epsilon -s) }{2} \right)
\pm\left(
\frac{(D-1)^2}{D} \epsilon +(k\pm \mu) D\right) \tan
\frac{\pi \alpha}{2}\\ \nonumber & \,&
\Phi_0(\epsilon)=\epsilon  \cot \left(\frac{\pi (\epsilon-s) }{2} \right)
+\mu\frac{D^2+ 2(D-1)^2}{D} \tan
\frac{\pi \alpha}{2}.
\end{eqnarray}

Let us consider the simplest case $ k=1  $, where $ \delta_0 =
\frac{1}{2} $, and  $ M_{eff}^2 = \tau_0 $.
In this case, it is clear from the above equation, that $
f(\epsilon)=0 $ unless
\begin{equation}
  \epsilon = \epsilon_0 - n \mu\\;\, n=0,1,\dots,
\end{equation}
where the highest power $ \epsilon_0 $ is determined by the equation
\begin{equation}
 \Phi_2(\epsilon_0)=0.
\end{equation}
This highest power term was already found in the previous paper

We solved this equation numerically in four dimensions and we found
the following
values of $ \epsilon_0 $ for two lowest values of $ \alpha^{\Lambda} $, in
spectroscopic notations $ \Lambda  = (-1)^s $
\begin{eqnarray}
  && 1.36901^+ :\epsilon_0 = \{2.08496, 4.11512 , 6.13072 ,
8.14028,\dots \},  \\ \nonumber & \,&
  1.36901^-:\epsilon_0=\{1.05590,3.10290,5.12399,7.13601,\dots\},
\\ \nonumber & \,&
  2.63099^+ :
\epsilon_0=\{1.94572,3.91608,5.89759,7.88500,\dots\},
\\ \nonumber &\,&
  2.63099^- : \epsilon_0=\{2.92897,4.90587,8.88011,10.87223,\dots\}.
\end{eqnarray}

Let us pick up a particular $ \epsilon_0 $, and let us study the
arising recurrent equation for the coefficients
\begin{equation}
  f(\epsilon)= \sum_{k=1}^{4} W_k(\epsilon) f(\epsilon+k\mu),
\end{equation}
where
\begin{eqnarray}
  &&W_1(\epsilon)=-\frac{(P^2+\tau_0) }{2A\Phi_2(\epsilon)}, \\
\nonumber & \,&
 W_2(\epsilon) = -\frac{2B\Phi_0(\epsilon)}{A\Phi_2(\epsilon)}, \\
\nonumber & \,&
W_3(\epsilon) = -\frac{B(P^2+\tau_0) }{2A^2\Phi_2(\epsilon)}, \\
\nonumber & \,&
 W_4(\epsilon) = -\frac{B^2\Phi_{-2}(\epsilon)}{A^2\Phi_2(\epsilon)}.
\end{eqnarray}
We could write down the formal solution
\begin{equation}
  f(\epsilon_0-n\mu) =\prod_{l=n}^{l=1} \left(\sum_{k=1}^{4}
W_k(\epsilon_0-l\mu)
\exp \left(k \frac{d}{d \epsilon_0} \right) \right)  f(\epsilon_0),
\end{equation}
where it is implied, that $ f(\epsilon) = 0$ at $ \epsilon >
\epsilon_0 $, and the operator ordering is as indicated, i.e., from $
l=n $ to $ l=1 $.

These coefficients should terminate at the smallest $
\epsilon > - \beta $, according to our boundary condition (\ref{BCON})
\begin{equation}
  f(\epsilon_0-n_0\mu)=0\\;\; n_0 =\lceil \frac{\epsilon_0+\beta}{\mu} \rceil.
\end{equation}
This provides us with the spectral equation, which is a polynomial in $ P^2
$.
The roots $ \xi_n $ of this polynomial correspond to the particle
spectrum. Restoring  quantum numbers,
\begin{equation}
  -P^2 = \tau_0 + \sqrt{AB} \,\xi_n(s,\epsilon_0).
\end{equation}

The corresponding roots  for the lowest levels
of $ \epsilon_0 $ are
\begin{eqnarray}
   &&1.36901^+, \epsilon_0 =2.08496:\xi_n=\pm\{
6.94993 \pm 25.62917\, \imath ,11.64012\pm 7.61785\,\imath\},
\\ \nonumber & \,&
   1.36901^-, \epsilon_0 =1.05590: \xi_n=\{0,0,0,0,0\},
\\ \nonumber & \,&
  2.63099^+, \epsilon_0=1.94572: \xi_n = 0,
\\ \nonumber & \,&
  2.63099^+,\epsilon_0=3.91608: \xi_n= \{0,0,0\}, \\
\nonumber & \,&
  2.63099^-,\epsilon_0=2.92897: \xi_n= \pm 49.97203, \\
\nonumber & \,&
  2.63099^-,\epsilon_0=4.90587: \xi_n= \{0,0,0\}.
\end{eqnarray}
With large enough $
\tau_0 $ there exist tachyon-free solutions for the second fixed
point, $ \alpha =2.63099 $,  but apparently, there are
no solutions with infinitely rising masses,
because of the sign degeneracy $ \xi \rightarrow -\xi $ in both
fixed points.

This is another indication of the triviality of the $k=1$
fixed points (they are the only ones with rational critical
index $ \delta_0= \frac{1}{2} $  of the mass spectrum). Most likely,
there is  the finite number of
the free scalar particles, i.e., these are just the Gaussian fixed points.

The case $ k>1 $ is  much more complicated, as there are
also negative integer powers of $ \lambda $ involved, apart from powers of $
\lambda^{-\mu} $ in the expansion. We are going to study this case in
the next paper.

To summarize, there are two branches of the spectrum. The odd
$ \lambda $-parity states  represent the scalar particles, dressed by
interaction
with the gauge fields. The even $ \lambda $-parity states represent the mixture
of "glueballs"
with the even number of scalar particles. The mass scale behaves as
certain irrational power of initial parameters of the lattice model.

In the strong coupling phase for the two simplest fixed points  with
rational  scaling indexes for masses, we computed the particle
spectrum. One fixed
point, with $ \rho(0)=0 $,  turned out unstable, and in the other one,
with $ \rho(0)= \infty $, there were stable solutions.
However, the spectrum did not rise to infinity, as one would expect in
QCD, which indicates triviality of these fixed points.
We leave for future study the exciting numerical problem of computing
masses from above equations of  higher
critical points in the strong coupling phase.

So far, we cannot even tell, whether the spectrum terminates, and
whether there are tachyons. Perhaps, some general inequalities can be
derived to answer this question.  One way to guarantee the absence of
tachyons in the given fixed point is is to arrive at this fixed point
from the lattice MFE, with real potential, stable at infinity. This
requires the serious numerical study of the MFE, and/or the
simulations of the initial lattice model.


\newsection{Acknowledgements}

I would like to thank the theory groups of Ecole Normale and
Jussieu in Paris for their hospitality, and David Gross, Volodja
Kazakov and Ivan Kostov for interesting discussions.
This work was partially supported by the National Science Foundation under
contract PHYS-90-21984.


\begin{thebibliography}{10}
\small
\addtolength{\itemsep}{-6pt}

\bibitem{KM92}
V.A.Kazakov and A.A.Migdal, {\it Induced QCD at large N}, Paris / Princeton
preprint LPTENS-92/15 / PUPT-1322 (June, 1992)

\bibitem{Mig92a}
A.A.Migdal, {\it Exact solution of induced lattice gauge theory at large N},
Princeton preprint PUPT-1323 (June, 1992)

\bibitem{Mig92b}
A.A.Migdal, {\it 1/N expansion and particle spectrum in induced QCD},
Princeton preprint PUPT-1332 (July, 1992)

\bibitem{GS92}
A.Gocksch and Y.Shen, {\it The phase diagram of the $N=2$ Kazakov-Migdal
model}, BNL preprint (July, 1992);

\bibitem{Gr92}
D.Gross, {\it Some remarks about induced QCD}, Princeton preprint PUPT-1335
(August, 1992)


\bibitem{CAP92}
M.Caselle, A.D.'Adda and S.Panzeri, {\it Exact solution of D=1
Kazakov-Migdal induced gauge theory}, Turin preprint DFTT 38/92 (July,
1992);


\bibitem{KhM81}
S.B.Khokhlachev and Yu.M.Makeenko, {\sl Phys.~Lett.} {\bf 101B} (1981) 403;
ZhETF {\bf 80} (1981) 448 (Sov.~Phys.~JETP {\bf 53} (1981) 228)

\bibitem{MC}
I.G.Holliday and A.Schwimmer, {\sl Phys.~Lett.} {\bf 101B} (1981) 327; \\
J.Greensite and B.Lautrup, {\sl Phys.~Rev.~Lett} {\bf47} (1981) 9; \\
G.Bhanot, {\sl Phys.~Lett.} {\bf 108B} (1982) 337; \\
M.Creutz and K.J.M.Moriarty, {\sl Nucl.~Phys.} {\bf B210[FS6]} 50

\bibitem{Sam83}
Yu.M.Makeenko and M.I.Polikarpov, {\sl Nucl.~Phys.} {\bf B205[FS5]} (1982) 386;
\\ S.Samuel, {\sl Phys.~Lett.} {\bf 112B} (1982) 237, {\bf 122B} (1983) 287

\bibitem{MF}
J.Greensite and B.Lautrup, {\sl Phys.~Lett.} {\bf 104B} (1981) 41; \\
P.Cvitanovi\'{c}, J.Greensite and B.Lautrup, {\sl Phys.~Lett.}
{\bf 105B} (1981) 197; \\
T.-L.Chen, C.-I~Tan and X.-T.Zheng, {\sl Phys.~Lett.} {\bf 109B} (1982) 383;
{\sl Phys.~Rev.} {\bf D26} (1982) 2843; \\
M.C.Ogilvie and A.Horowitz, {\sl Nucl.Phys.} {\bf B215} (1983) 249

\bibitem{Monopole}
I.G.Holliday and A.Schwimmer, {\sl Phys.~Lett.} {\bf 102B} (1981) 337; \\
R.C.Brower, D.A.Kessler and H.Levine, {\sl Nucl.~Phys.} {\bf B205[FS5]} (1982)
77; \\
L.Caneschi, I.G.Holliday and A.Schwimmer, {\sl Nucl.~Phys.} {\bf B200[FS4]}
(1982)
409

\bibitem{KSW92}
I.I.Kogan, G.W.Semenoff and N.Weiss, {\it Induced QCD and hidden local $Z_N$
symmetry}, UBC preprint UBCTP-92-022 (June, 1992)


\bibitem{KhM92}
S.B.Khokhlachev and Yu.M.Makeenko,{\it The problem of large-N phase
transition in Kazakov-Migdal model of induced QCD }, ITEP-YM-5-92,
(July, 1992)

\bibitem{KMSW92}
I.I.Kogan, A.Morozov, G.W.Semenoff and N.Weiss, {\it Area law and
continuum limit in "induced QCD"}, UBC preprint UBCTP-92-022 (June, 1992)

\end{thebibliography}
\end{document}